\documentclass[12pt]{article}
\usepackage{amsfonts}

\input{tcilatex}

\begin{document}

\begin{center}
\textbf{PROPAGATION AND INTERACTION OF ULTRASHORT ELECTROMAGNETIC PULSES IN
NONLINEAR MEDIA WITH A QUADRATIC-CUBIC NONLINEARITY}

\vspace{0.5in}Elena V. Kazantseva$^{a}$\footnote{%
electronic addres: e-lena@pico.mephi.ru} and Andrey I. Maimistov$^{a}$ 
\footnote{%
electronic address: maimistov@pico.mephi.ru}

Boris A. Malomed$^{b}$\footnote{%
electronic address malomed@eng.tau.ac.il}

\vspace{0.2cm}$^a$Department of Solid State Physics, Moscow Engineering
Physics Institute, Moscow, 115409, Russia

$^{b}$Department of Interdisciplinary Studies, Faculty of Engineering, Tel
Aviv University, Tel Aviv 69978, Israel
\end{center}

\newpage

\begin{center}
\textbf{ABSTRACT}
\end{center}

\bigskip

Propagation of extremely short unipolar pulses of electromagnetic field
(``videopulses'') is considered in the framework of a model in which the
material medium is represented by anharmonic oscillators (approximating
bound electrons) with quadratic and cubic nonlinearities. Two families of
exact analytical solutions (with positive or negative \textit{polarity}) are
found for the moving solitary pulses. Direct simulations demonstrate that
the pulses are very robust against perturbations. Two unipolar pulses
collide nearly elastically, while collisions between pulses with opposite
polarities and a small relative velocity are inelastic, leading to emission
of radiation and generation of a small-amplitude additional pulse.


\newpage

\section{INTRODUCTION}

Very short nonlinear pulses of the electromagnetic field, which contain few
optical cycles \cite{DS,Kaplan,Kaplan2,Sazonov,Sazonov2}, or even of half a
cycle \cite{Kaplan3}, have recently attracted a great deal of attention.
They are referred to as ultrashort pulses (USP), or \textit{videopulses} 
\cite{DS,Sazonov, Sazonov2}. From the viewpoint of applications, the
interest to these pulses is stipulated by a perspective of a drastic
increase in the rates of data transmission and procession. One of
possibilities for generation of USPs is compression of a preliminarily
phase-modulated pulse. The phase modulation can be induced by the Kerr
(cubic) nonlinearity \cite{Agrawal}. For example, pulses with the duration $%
\tau _{p}=4.5$ fs, consisting of two optical cycles with the frequencies
belonging to the visible range, and with $\tau _{p}=40$ fs, containing just
one optical cycle at the frequency of the CO$_{2}$ laser, were reported in
Ref. \cite{Babzien}.

If the width of a pulse is much larger than the period of the optical cycle,
the pulse can be naturally described by means of its slowly varying
envelope, whose evolution is governed by an effective parabolic (nonlinear
Schr\"{o}dinger) equation \cite{Agrawal}. However, this approximation cannot
be applied to USPs, which calls for development of other theoretical models
admitting a relatively simple description of \ the propagation of extremely
short pulses.

The existing models that can be used for the description of USPs fall into
two classes, resonant and nonresonant ones. A resonant medium is modeled by
an ensemble of atoms with discrete energy levels. The field frequency being
close to a frequency of an atomic transition, only the (nearly) resonant
component of the field\ is taken into account. In particular, the simplest
model of the \textit{self-induced transparency} assumes a medium consisting
of two-level atoms resonantly coupled to the field. In the case of
multi-level atoms, it may be necessary to consider several monochromatic
components of the field with different frequencies.

In the works \cite{Bullough, Akimoto, Andreev}, USP solutions to a closed
system of the Maxwell-Bloch equations have been found analytically without
assuming a separation into the carrier wave and envelope. Generally, the
Maxwell's equations admit the propagation of electromagnetic waves in both
directions. If, however, the nonlinear contribution into the medium's
polarization is small, the \textit{unidirectional} wave propagation may be
assumed. This approximation admits reducing the wave equation to the
first-order one without any assumption about the shape of the waves.

The nonlinear dynamics of the medium driven by the electromagnetic filed is
frequently modeled in terms of anharmonic oscillators. In particular, the
propagation of a linearly polarized USP was considered in Refs. \cite
{Kaplan, ME92} in the framework of the material model based on the Duffing
oscillators, the nonlinear response of the medium being cubic. The
propagation of a linearly polarized USP in a dispersive medium was modeled
by means of quadratically nonlinear oscillators in Ref. \cite{DS} (without
dispersion, this was done in Ref.\cite{Akopyan}).

In all these works, the oscillator represents a response of the
high-frequency electron degree of freedom to the electromagnetic field.
Since the USP spectrum is concentrated at low frequencies, ion oscillations
may also give a considerable contribution to the full polarization of the
medium. Nevertheless, we will neglect the ion component in the material
response (the propagation of femtosecond pulses in a medium with the
nonlinearity determined by both electronic and ionic (Raman-scattering)
degrees of freedom was considered in Ref. \cite{Kozlov}).

An objective of the present work is to study the \emph{unidirectional}
propagation and interactions of linearly polarized USPs in a nonlinear
dispersive medium modeled by an anharmonic oscillator combining quadratic
and cubic nonlinearities. As is well known, in the case when the oscillator
is quasi-harmonic, the quadratic and cubic nonlinear terms produce effects
of the same order of magnitude \cite{LL}, that is why it is natural to
consider a mixed model of this type. It is also relevant to mention that
modeling dynamics of the broad (rather than ultrashort) optical solitons in
a medium with competing quadratic and cubic nonlinearities has recently
attracted considerable attention, see, e.g., the works \cite{quadrcubic} and
references therein.

The rest of the paper is structured as follows. The model is derived in
section 2. Two families of moving USP solutions with positive and negative
polarities are found analytically in section 3 (only one family survives in
the limit of a vanishing cubic nonlinearity). Stability of the pulses and
their collisions (for both signs of the relative polarity of the colliding
pulses) are investigated in section 4 by means of direct simulations.

\section{THE\ MODEL}

The one-dimensional propagation of the electromagnetic waves in a nonlinear
medium is governed by the wave equation,

\begin{equation}
\frac{\partial ^{2}E}{\partial z^{2}}-\frac{1}{c^{2}}\frac{\partial ^{2}E}{%
\partial t^{2}}=\frac{4\pi }{c^{2}}\frac{\partial ^{2}P}{\partial t^{2}},
\label{wave}
\end{equation}
where $P$ is the polarization of the medium. In the framework of the usual
unidirectional approximation \cite{BJKS, Eilb}, Eq. (\ref{wave}) can be
reduced to the first-order equation,

\begin{equation}
\frac{\partial E}{\partial z}+\frac{1}{c}\frac{\partial E}{\partial t}=- 
\frac{2\pi }{c}\frac{\partial P}{\partial t}.  \label{eq1}
\end{equation}

We adopt a simple anharmonic-oscillator model for the medium, which is a
frequently used approximation \cite{Bloemb} (see also \cite{Yariv, Butch}).
If $X$ as the displacement of an electron from its equilibrium position, the
motion equation (which neglects friction) can be written as

\begin{equation}
\frac{\partial ^{2}X}{\partial t^{2}}+\omega _{0}^{2}X-\kappa
_{2}X^{2}+\kappa _{3}X^{3}=\frak{L}\frac{e}{m}E(z,t),  \label{eq2}
\end{equation}
where $\omega _{0}$ is an eigenfrequency of the oscillator, while $\kappa
_{2}$ and $\kappa _{3}$ are anharmonicity coefficients. The term on the
right-hand side of Eq. (\ref{eq2}) represents the force exerted on the
electron by the electromagnetic field, where $\frak{L}=(\varepsilon +2)/3$
is the Lorentz factor, $e$ is the electron's electric charge, and $m$ is its
mass. However, it is possible to absorb $\frak{L}$ into an effective mass $%
m_{\mathrm{eff}}\equiv m/\frak{L}$ . Hereafter, we will use $m$ as a symbol
for this effective mass. Finally, the dynamical variable $X$ is related to
the medium's polarization, $P=n_{A}eX$, where $n_{A}$ is the density of the
oscillators (atoms).

We define rescaled variables, 
\begin{equation}
\zeta \equiv z/l,\tau \equiv \omega _{0}(t-z/c),\,\mathcal{E}\equiv
E/A_{0},\,\mathrm{and\,\,\,}q\equiv X/X_{0},  \label{rescaling}
\end{equation}
where $A_{0}=m\omega _{0}^{4}/e|\kappa _{2}|,\quad X_{0}=\omega
_{0}^{2}/|\kappa _{2}|,\quad 1/l=2\pi n_{A}e^{2}/(mc\omega _{0})=\omega
_{p}^{2}/2\omega _{0}c$, and 
\begin{equation}
\omega _{p}=(4\pi n_{A}e^{2}/m)^{-1/2}  \label{omega_p}
\end{equation}
is the plasma frequency. In terms of the rescaled variables, Eqs. (\ref{eq1}%
) and (\ref{eq2}) take the form

\begin{equation}
\frac{\partial \mathcal{E}}{\partial \zeta }=-\frac{\partial q}{\partial
\tau },\quad \frac{\partial ^{2}q}{\partial \tau ^{2}}+q-q^{2}+2\mu q^{3}=%
\mathcal{E}\,,  \label{eq3}
\end{equation}
with the single remaining parameter $\mu =(\kappa _{3}\omega
_{0}^{2}/2\kappa _{2}^{2})$. Equations (\ref{eq3}) furnish a final form of
the model.

To estimate the magnitude of parameter $\mu $, we consider Morse potential

\[
U_{M}(X)=U_{0}[\exp (-2X/r_{0})-2\exp (-X/r_{0})],
\]
where $r_{0}$\ is an atomic characteristic size and $U_{0}$\ is a ionization
potential. The expansion of this potential in terms of a power series in $X$%
\ \ leads to $\omega _{0}^{2}=2U_{0}/mr_{0}^{2},|\kappa
_{2}|=3U_{0}/mr_{0}^{3},\kappa _{3}=7U_{0}/3mr_{0}^{4}.$ Thus $\mu
=7/27\approx 0.259$.

Consider now Kepler's potential with Coulomb long-range attraction:

\[
U_{K}(X)=U_{0}\left[ \left( \frac{r_{0}}{r_{0}+X}\right) ^{2}-2\left( \frac{%
r_{0}}{r_{0}+X}\right) \right] , 
\]
where $r_{0}$ corresponds to equilibrium position. In this case we have $%
\omega _{0}^{2}=2U_{0}/mr_{0}^{2}, |\kappa _{2}|=6U_{0}/mr_{0}^{3}, \kappa
_{3}=12U_{0}/mr_{0}^{4}$ ,\ and $\mu =1/3$. In following numerical
simulation of the steady state pulses propagation about the same magnitude
of parameter $\mu $ \ will be used.

\section{DYNAMICAL\ INVARIANTS}

It is worthy to note that the system of equations (\ref{eq3}) can be derived
as the Euler-Lagrange equations from the action functional

\[
S=\int \mathcal{L}[q,\phi ]dxdt, 
\]
where the Lagrangian density is

\begin{equation}
\mathcal{L}=\frac{1}{2}\frac{\partial \phi }{\partial x}\frac{\partial \phi 
}{\partial t}+\frac{1}{2}\left( \frac{\partial q}{\partial x}\right) ^{2}-%
\frac{1}{2}q^{2}+\frac{1}{2}q^{3}-\frac{\mu }{2}q^{4}+q\frac{\partial \phi }{%
\partial x},  \label{Lagrangi}
\end{equation}
and the independent variables were temporarily (up to the end of this
section) redenoted as $\zeta \rightarrow t,\tau \rightarrow x$, in order to
have the notation similar to that in the classical field theory (a
definition of $\phi $ is given below).

Application of the variational procedure to the action $S$ yields equations

\begin{equation}
\frac{\partial ^{2}\phi }{\partial t\partial x}+\frac{\partial ^{2}q}{%
\partial x^{2}}=0,\quad \frac{\partial ^{2}q}{\partial t^{2}}+q-q^{2}+2\mu
q^{3}=\frac{\partial \phi }{\partial x}.  \label{eqL9}
\end{equation}
Identifying $\phi $ as a potential for the fields $q$ and $\mathcal{E}$, so
that $q\equiv -\partial \phi /\partial t$ and $\mathcal{E}\equiv \partial
\phi /\partial x$, makes these equations identical to Eqs. (\ref{eq3}),
which can be further transformed into a single equation

\begin{equation}
\frac{\partial q}{\partial t}+\frac{\partial q}{\partial x}-2q\frac{\partial
q}{\partial t}+6\mu q^{2}\frac{\partial q}{\partial t}+\frac{\partial ^{3}q}{%
\partial t\partial x^{2}}=0.  \label{eqL10}
\end{equation}

Note that Eq. (\ref{eqL10}) may be regarded as a continuity equation,

\begin{equation}
\frac{\partial }{\partial t}\left( q-q^{2}+2\mu q^{3}+\frac{\partial ^{2}q}{%
\partial x^{2}}\right) +\frac{\partial q}{\partial x}=0.  \label{eqL14}
\end{equation}
Integration of Eq. (\ref{eqL14}) in $x$ leads to

\[
\frac{\partial }{\partial t}\int\limits_{-\infty }^{+\infty }\left(
q-q^{2}+2\mu q^{3}\right) dx+\frac{\partial }{\partial t}\left( \frac{%
\partial q}{\partial x}|_{{-\infty }}^{+\infty }\right) +q|_{-\infty
}^{+\infty }=0. 
\]
Assuming zero boundary conditions for $q$ and its derivatives at $%
x\rightarrow \pm \infty $, the latter relation implies the conservation of a
dynamical invariant,

\begin{equation}
I_{1}=\int\limits_{-\infty }^{+\infty }\left( q-q^{2}+2\mu q^{3}\right) dx.
\label{eqL15}
\end{equation}
By taking into account the second equation from Eqs. (\ref{eq3}) in the form

\begin{equation}
\frac{\partial ^{2}q}{\partial x^{2}}+q-q^{2}+2\mu q^{3}=\mathcal{E},
\label{prom1}
\end{equation}
one can find that $I_{1}$ is the ``area of the pulse'',

\begin{equation}
I_{1}=\int_{-\infty }^{+\infty }\mathcal{E\,}dx.  \label{eqL16}
\end{equation}

To find another dynamical invariant, we multiply Eq. (\ref{eqL10}) by $q$
and transform the resulting relation into a form

\begin{equation}
\frac{\partial }{\partial t}\left[ \frac{1}{2}q^{2}-\frac{2}{3}q^{3}+\frac{
3\mu }{2}q^{4}-\frac{1}{2}\left( \frac{\partial q}{\partial x}\right) ^{2} %
\right] +\frac{\partial }{\partial x}\left( \frac{1}{2}q^{2}+q\frac{\partial
^{2}q}{\partial x\partial t}\right) =0.  \label{eqL17}
\end{equation}
Thus, a new continuity equation is obtained, giving rise to a new dynamical
invariant,

\begin{equation}
I_{2}=\int\limits_{-\infty }^{+\infty }\left[ \frac{1}{2}q^{2}-\frac{2}{3}
q^{3}+\frac{3\mu }{2}q^{4}-\frac{1}{2}\left( \frac{\partial q}{\partial x}
\right) ^{2}\right] \,dx.  \label{eqL18}
\end{equation}

Let us now get back to the conserved density in the continuity equation (\ref
{eqL14}),

\[
Q\equiv q-q^{2}+2\mu q^{3}+\frac{\partial ^{2}q}{\partial x^{2}}. 
\]
Multiplying both sides of Eq. (\ref{eqL14}) by $Q$, we obtain

\[
\frac{\partial }{\partial t}\left( \frac{1}{2}Q^{2}\right) +Q\frac{\partial q%
}{\partial x}=0. 
\]
This immediately leads to a continuity equation

\begin{equation}
\frac{\partial }{\partial t}\left( \frac{1}{2}Q^{2}\right) +\frac{\partial }{%
\partial x}\left[ \frac{1}{2}q^{2}-\frac{1}{3}q^{3}+\frac{\mu }{2}q^{4}+ 
\frac{1}{2}\left( \frac{\partial q}{\partial x}\right) ^{2}\right] =0
\label{eqL19}
\end{equation}
and the third dynamical invariant,

\begin{equation}
I_{3}=\frac{1}{2}\int\limits_{-\infty }^{+\infty }\left( q^{2}-q^{3}+2\mu
q^{4}+\frac{\partial ^{2}q}{\partial x^{2}}\right) ^{2}dx.  \label{eqL20a}
\end{equation}
Taking into account the relation (\ref{prom1}), this integral may be
interpreted as a ``pulse energy'',

\begin{equation}
I_3=\frac 12\int _{-\infty }^{+\infty }\mathcal{E}^2dx.  \label{eqL20b}
\end{equation}

To find the interpretation of the dynamical invariant given by Eq. (\ref
{eqL18}), we resort to the Lagrangian density (\ref{Lagrangi}). The density
of the \emph{canonical Hamiltonian} for this dynamical system can be
obtained from $\mathcal{L}$ by means of the standard \textit{Legendre
transformation,}

\[
\mathcal{H}=\frac{\partial \mathcal{L}}{\partial \phi _{,t}}\phi _{,t}-\frac{%
\partial \mathcal{L}}{\partial q_{,t}}q_{,t}-\mathcal{L}=-\frac{1}{2}\left( 
\frac{\partial q}{\partial x}\right) ^{2}+\frac{1}{2}q^{2}-\frac{1}{3}q^{3}+%
\frac{\mu }{2}q^{4}-q\mathcal{E}. 
\]
The variable $\mathcal{E}$ can be eliminated from it, using Eq. (\ref{prom1}%
), so that

\begin{equation}
\mathcal{H}=-\frac{\partial }{\partial x}\left( q\frac{\partial q}{\partial x%
}\right) +\frac{1}{2}\left( \frac{\partial q}{\partial x}\right) ^{2}-\frac{1%
}{2}q^{2}+\frac{2}{3}q^{3}-\frac{3\mu }{2}q^{4},  \label{Hdensity}
\end{equation}
Omitting the full derivative, the Hamiltonian corresponding to the density (%
\ref{Hdensity}) takes the form

\[
H=\int\limits_{-\infty }^{+\infty }\left[ \frac{1}{2}\left( \frac{\partial q%
}{\partial x}\right) ^{2}-\frac{1}{2}q^{2}+\frac{2}{3}q^{3}-\frac{3\mu }{2}
q^{4}\right] \,dx\equiv -I_{2}\,, 
\]
cf. Eq. (\ref{eqL18}). Thus, the dynamical invariant $-I_{2}$ is nothing
else but the canonical Hamiltonian of the system under consideration.

\section{ANALYTICAL SOLUTIONS FOR THE ULTRASHORT PULSES}

It seems plausible that the system of Eqs. (\ref{eq3}) is not an integrable
one. Nevertheless, some exact analytical solutions, describing the
propagation of USPs, can be found. To this end, one assumes that $\mathcal{E}
$ and $q$ depend on a single variable, 
\begin{equation}
\eta \equiv \tau -\zeta /\alpha =\omega _{0}(t-z/V),  \label{eta}
\end{equation}
with some constant $\alpha $. An expression for the velocity $V$ of a
steadily moving pulse then follows from Eq. (\ref{rescaling}),

\begin{equation}
\frac{1}{V}=\frac{1}{c}\left[ 1+\frac{1}{2\alpha }\left( \frac{\ \omega _{p}%
}{\omega _{0}}\right) ^{2}\right] \,,  \label{eq4}
\end{equation}
$\omega _{p}$ being the plasma frequency defined by Eq\textbf{. (}\ref
{omega_p}\textbf{).} The first equation of the system (\ref{eq3}) can be
integrated to yield

\begin{equation}
\mathcal{E}=\alpha q.  \label{eq5}
\end{equation}
Next, the second equation from the system (\ref{eq3}) takes the form

\begin{equation}
\frac{d^{2}q}{d\eta ^{2}}-(\alpha -1)q-q^{2}+2\mu q^{3}=0,  \label{eq6}
\end{equation}
This equation can be integrated once,

\begin{equation}
\left( \frac{dq}{d\eta }\right) ^{2}-(\alpha -1)q^{2}-\frac{2}{3}q^{3}+\mu
q^{4}=\mathrm{const},  \label{eq7}
\end{equation}
As we are interested in solitary-wave solutions, it is necessary to set $%
\mathrm{const}\,=\,0$, so that (\ref{eq7}) yields

\begin{equation}
\frac{dq}{d\eta }=\pm q\sqrt{(\alpha -1)+\frac{2}{3}q-\mu q^{2}},
\label{eq9}
\end{equation}

A substitution $q\equiv 1/y$ \ transforms Eq. (\ref{eq9}) into

\begin{equation}
\frac{dy}{d\eta }=\pm \sqrt{(\alpha -1)\left[ \left( y+\frac{1}{3(\alpha -1)}%
\right) ^{2}-\frac{1+9\mu (\alpha -1)}{9(\alpha -1)^{2}}\right] }\,.
\label{eq11}
\end{equation}
A new dependent variable $\xi $ will be used, defined by 
\[
y+\frac{1}{3(\alpha -1)}\equiv \sigma \left( \frac{1+9\mu (\alpha -1)}{%
9(\alpha -1)^{2}}\right) ^{1/2}\xi (\eta )\,, 
\]
with $\sigma =\pm 1$. In terms of $\xi $, Eq. (\ref{eq11}) is

\[
\sigma \frac{d\xi }{d\eta }=\pm \sqrt{(\alpha -1)\left( \xi ^{2}-1\right) }%
\,, 
\]
from where one finds $\xi (\eta )=\cosh (\sqrt{(\alpha -1)}\eta )$.

Thus, we have obtained a family of exact solutions parameterized by the
continuous \emph{positive} parameter $(\alpha -1)$ and discrete one $\sigma
=\pm 1$,

\begin{equation}
q^{(+)}(\eta ;\alpha )=\frac{3(\alpha -1)}{\sqrt{1+9(\alpha -1)\mu }\cosh (%
\sqrt{(\alpha -1)}\eta )-1},  \label{eq13.1}
\end{equation}
\begin{equation}
q^{(-)}(\eta ;\alpha )=-\,\frac{3(\alpha -1)}{\sqrt{1+9(\alpha -1)\mu }\cosh
(\sqrt{(\alpha -1)}\eta )+1}\,,  \label{eq13.2}
\end{equation}
the superscript standing for $\sigma $.

The pulses represented by the solutions (\ref{eq13.1}) and (\ref{eq13.2})
have different \textit{polarities}. In the limit $\mu \rightarrow 0$,
corresponding to the model with a purely quadratic nonlinearity, Eq. (\ref
{eq13.1}) goes over into a singular solution, 
\[
q_{\mathrm{quadr}}^{(+)}(\alpha ;\eta )=\frac{3(\alpha -1)}{2\sinh ^{2}(%
\sqrt{(\alpha -1)}\eta /2)}\,, 
\]
while Eq. (\ref{eq13.2}) yields a nonsingular pulse in the same limit, which
was already found in Ref. \cite{Maimistov}, 
\[
q_{\mathrm{quadr}}^{(-)}(\alpha ;\eta )=-\,\frac{3(\alpha -1)}{2\cosh ^{2}(%
\sqrt{(\alpha -1)}\eta /2)}\,\,. 
\]
Thus, in compliance with the results of Ref. \cite{Maimistov}, only one
family of nonsingular pulses exists in the case of the purely quadratic
nonlinearity.

By using Eq. (\ref{eq5}), we retrieve an expression for the electromagnetic
field corresponding to the solutions (\ref{eq13.1}) and (\ref{eq13.2}),

\begin{equation}
\mathcal{E}^{(+)}(\alpha ;\eta )=\frac{3\alpha (\alpha -1)}{\sqrt{1+9(\alpha
-1)\mu }\cosh (\sqrt{(\alpha -1)}\eta )-1},  \label{eq14.1}
\end{equation}

\begin{equation}
\mathcal{E}^{(-)}(\alpha ;\eta )=-\,\frac{3\alpha (\alpha -1)}{\sqrt{%
1+9(\alpha -1)\mu }\cosh (\sqrt{(\alpha -1)}\eta )+1}.  \label{eq14.2}
\end{equation}

We stress that the velocities of two pulses which have the opposite
polarities but the same value of $(\alpha -1)$ are equal, since they are
determined by Eq. (\ref{eq4}) and depend only on $\alpha $.

\section{NUMERICAL SIMULATION OF PROPAGATION\ AND\ COLLISIONS\ OF\ THE\
PULSES}

Although the pulse solutions have been obtained in the exact form, their
stability and interactions should be studied by means of numerical
simulations. To simulate the system (\ref{eq3}), it is convenient to
transform it into the following form:

\begin{equation}
\frac{\partial \mathcal{E}}{\partial \zeta }=-p,  \label{eq15.1}
\end{equation}

\begin{equation}
\frac{\partial q}{\partial \tau }=p,\quad \frac{\partial p}{\partial \tau }=%
\mathcal{E}-q+q^{2}-2\mu q^{3}.  \label{eq15.2}
\end{equation}
Since Eq. (\ref{eq15.1}) contains only the derivative with respect to $\zeta 
$, and the derivatives with respect to $\tau $ are present only in Eqs. (\ref
{eq15.2}), one can use any technique of numerical integration of ordinary
differential equations. For solving Eqs. (\ref{eq15.2}), we employed the
prediction-correction method. The fourth-order Runge-Kutta routine was used
for integrating Eq. (\ref{eq15.1}). As the initial conditions, we took the
analytical solutions given by Eqs. (\ref{eq13.1}) and (\ref{eq14.1}), or (%
\ref{eq13.2}) and (\ref{eq14.2}), at $\zeta =0$ (with regard to Eq. (\ref
{eta})). The boundary conditions are $q(\zeta ,\tau )=p(\zeta ,\tau )=0$ at $%
|\tau |\longrightarrow \infty $.

By using the Morse $U_{M}(X)$ and Kepler $U_{K}(X)$ \ potentials one can
estimate the normalized amplitudes $A_{0}$ and $X_{0}$. For Morse potential $%
A_{0}=(4/3)(U_{0}/er_{0}),X_{0}=(2/3)r_{0},$ and for Kepler's potential we
have $A_{0}=(2/3)(U_{0}/er_{0}),X_{0}=(1/3)r_{0}$. In both case $%
(U_{0}/er_{0})$ can be identified with atomic field \cite{Butch}.
Consequently, the peak amplitude of the steady state pulse must be not
exceed $(U_{0}/er_{0}).$ Furthermore, the anharmonic-oscillator model
proposes that the magnitude of $q$ is less than one. It leads to the
inequality $(\alpha -1)<1$. However, in order to obtain the descriptive
illustrations of the numerical results the parameter $\alpha $ have been
taken both from the interval $1.01<\alpha <1.1$ and from the interval $%
2<\alpha <8.$

A typical numerical result, with $\alpha =7.25$ and $\mu =2/9$, is displayed
in Fig.1, which demonstrates stable propagation of the pulse, at least up to 
$\zeta =200$. To further test the robustness of USPs, we added to the
initial configuration perturbations in the form of a quasi-harmonic or
biharmonic wave with amplitudes $\leq 0.1$ of the pulse's amplitude, and
with different frequencies. As a typical example, Fig.~2 demonstrates
evolution of the pulse with $\alpha =5$ and $\mu =1/3$ , with the initially
added perturbation in the form of a long quasi-monochromatic packet filled
by the wave $\delta \mathcal{E}(\tau )=\delta \mathcal{E}_{0}\sin (15\tau )$%
. As is seen in from Fig.~2, USP is not scathed by the perturbation. With
the increase of the amplitude of the quasi-harmonic perturbation, we
observed no conspicuous change of USP. Simulations of the pulses evolution
in the presence of a biharmonic (two-frequency) perturbation (not shown
here) showed that the increase of the amplitudes of the biharmonic
perturbation results in a small change of the value of the parameter $\alpha 
$ corresponding to the finally established USP. A similar effect was
generated by an initial perturbation in the form of an additional
rectangular pulse superimposed on USP, see an example in Fig.~3. The
increase of the perturbation's amplitude leads to an increase of the final
value of the USP's amplitude and decrease of its velocity.

Simulations of interactions between pulses with widely different values of
the parameter $\alpha $ (i.e., velocities) have shown that their collisions
are quasi-elastic, irrespective of the polarities of the colliding pulses:
after passing through each other, the pulses retrieve the same shapes and
velocities as they had before the collision, the only result being a shift
of their centers. However, the character of the interaction between pulses
with equal and opposite polarities becomes different with the decrease of
their relative velocity. In the case of equal polarities, the pulses with a
small relative velocity demonstrate strong mutual repulsion: the distance
between them attains a minimum, and then they start to separate again, so
that they never completely overlap. Strong energy exchange between the
pulses takes place around the point where they attain the minimum
separation. The energy exchange gives rise to mutual interconversion of the
two pulses, so that after the collision they, effectively, swap their
positions. This picture, a typical example of which is displayed in Fig. 4,
is quite similar to the classical description of collisions between solitons
in the Korteweg - de Vries equation \cite{Ablo}.

Collisions between pulses with opposite polarities and close velocities are
found to be more inelastic in comparison with the unipolar pulses. In this
case, dispersive wave packets with a considerable amplitude are generated,
see an example in Fig. 5. Moreover, a new small-amplitude pulse with the
negative polarity is also generated by the inelastic collision shown in Fig.
5.

Another noteworthy feature revealed by the simulations is that the results
of the collisions are very different depending on which solitons (faster or
slower ones) have initially positive and negative polarities. This feature
is obvious from the comparison of Figs. 5 and 6, which differ by the
polarity reversal of the initial state. As is seen, in the latter case the
inelasticity is much weaker, and, in particular, no additional pulse is
generated.

To further illustrate the inelasticity of the collisions, in Fig. 7 we
display the shapes of the fields before and after collisions that were shown
in Figs. 5 and 6. In the former case (Fig. 7a), the amplitudes of the
positive- and negative-polarity pulses decreases and increases,
respectively, as a result of the collision. In the latter case (Fig. 7b),
the result is the opposite. From here, we conclude that, in the course of
the collision, the energy is always transferred to a more powerful pulse
(the one with a larger value of $\alpha $).

We also explored interactions between two pulses with equal values of $%
\alpha $ (hence, they have zero relative velocity), initially placed at
various distances from each other. In Fig.8, solid lines show trajectories
of the motion of centers of the interacting pulses with different
polarities, and dotted lines show the same for interacting unipolar pulses.
In the case shown in this figure, the initial distance between the pulses is
three half-widths of a steady-state pulse. At such a distance, tails of the
pulses overlap considerably, giving rise to energy transfer from one pulse
to the other (irrespective of the polarities, no visible interaction between
pulses takes place if the initial distances is increased to five
half-widths). As a result, the pulses cease to be identical and start to
separate (the pulse with a larger amplitude is moving slower). Eventually,
the pulses assume permanent (but different) shapes after the separation.

In the case of two unipolar pulses, energy is gained by the rear one,
therefore the trajectories of the pulses do not intersect in this case, see
Fig. 8. On the contrary, in the case of opposite polarities the front
positive-polarity pulse gains energy as a result of the interaction, while
the absolute value of the amplitude of the negative-polarity rear pulse
decreases and its velocity increases, hence the trajectories of the two
pulses intersect. After their collision, there appears a new small-amplitude
pulse with negative polarity (its trajectory is not shown in Fig. 8), cf.
Figs. 5 and 7a.

We stress that we have \emph{never} observed formation of a bound state of
pulses with equal or opposite polarities as a result of their interaction,
nor was it possible to find bound states in any other way.

To conclude this section, it is relevant to compare the present model with
some others. Indeed, after transforming the system into the single equation (%
\ref{eqL10}), replacing $\zeta $ and $\tau $ by $t$ and $x$, $q$ by $-u$,
and setting $\mu =0$ one obtains

\begin{equation}
\frac{\partial u}{\partial t}+\frac{\partial u}{\partial x}+2u\frac{\partial
u}{\partial t}+\frac{\partial ^{3}u}{\partial t\partial x^{2}}=0,
\label{eq18}
\end{equation}
which resembles to the so-called regularized long-wave equation \cite
{Courtency, Santarelli},

\begin{equation}
\frac{\partial u}{\partial t}+\frac{\partial u}{\partial x}+2u\frac{\partial
u}{\partial x}-\frac{\partial ^{3}u}{\partial t\partial x^{2}}=0.
\label{eq19}
\end{equation}
While Eqs. (\ref{eq18}) and (\ref{eq19}) have similar traveling-pulse
solutions, the equations are \emph{not} equivalent, hence they may produce
very different dynamical effects.

\section{CONCLUSION}

We have introduced and analyzed a model for the propagation of ultrashort
unipolar pulses of electromagnetic field in a material medium represented by
anharmonic oscillators with quadratic and cubic nonlinearities. Two families
of exact analytical solutions, with positive and negative polarities, have
been found for moving solitary pulses. Direct simulations have demonstrated
strong stability of the pulses against various perturbations. Collisions
between the pulses were also simulated in detail, showing that they interact
nearly elastically, irrespective of the pulses' relative polarity, unless
their relative velocity is very small. If the case of the small relative
velocity, collisions are inelastic, resulting in generation of radiation and
a new small-amplitude soliton.

\section*{Acknowledgment}

We are grateful to S.V. Sazonov and S.A.Kozlov for valuable discussions. A
part of this work was supported by INTAS (European Union) under the grant
No. 96-0339.

\newpage

\section{FIGURE CAPTIONS}

Fig.~1. Numerically simulated propagation of the positive-polarity (a) and
negative-polarity (b) ultrashort pulses predicted by the analytical
solutions (\ref{eq13.1}), (\ref{eq14.1}) and (\ref{eq13.2}), (\ref{eq14.2})
with $\alpha =7.25$ and $\mu =2/9$.

Fig.~2. Numerically simulated evolution of the pulse with $\alpha =5$ and $%
\mu =1/3$, with a superimposed perturbation in the form of a quasi-harmonic
wave packet.

Fig.~3. The same as in Fig.~2, with an initial perturbation in the form of
an additional rectangular pulse.

Fig.~4. An example of a numerically simulated collision between two unipolar
pulses with close velocities, corresponding to $\alpha =5$ and $\alpha =4.61$%
, at $\mu =1/3$.

Fig.~5. Collision between pulses with the negative and positive polarities
in the case of $\mu =1/3$. The pulse of negative polarity relates to $\alpha
=5$ and pulse with positive polarity corresponds with $\alpha =3.89$ .

Fig.~6. Collision of pulses with the same values of $\alpha $ and $\mu $\ as
in a fig. 5, but with reversed polarities.

Fig.~7. The shapes of the pulses before and after the collisions, the panels
(a) and (b) pertaining to the two cases shown in Figs. 5 and 6.

Fig.~8. Trajectories of the motion of two interacting soliton with exactly
equal initial velocities (equal values of $\alpha $). The solid and dotted
pairs of lines pertain, respectively, to the solitons with opposite and
equal polarities.

\end{document}